{\overfullrule=0mm}
\parindent=0mm

\def\Thm#1{{{\bf Theorem 1.}\sl#1}\rm}
\def\Thmm#1{{{\bf Theorem 2.}\sl#1}\rm}

\def\Pr{\bf Proof.\rm}
\def\Rp{\hbox{\carre\rm}}
\def\Def#1{{{\bf Definition.}\sl#1}\rm}
\def\bn{\bigskip\nobreak}
\def\bn{\sk}

\def\ttparag#1{\sk{\bfXII #1}\sk\sk} 
\hsize=15truecm
\vsize=20truecm
\voffset=0cm
\hoffset=0mm
\newtoks\gauche
\newtoks\droite
\def\makeheadline{\vbox to 0pt{\vskip -40pt\line{\vbox to 8.5pt{}\the
\headline}\vss\nointerlineskip}}
\headline={{\voffset 2cm\sevenbf\the\gauche\hfill\the\droite}}
\def\sk{\smallskip}

\font\XII=cmr10 scaled\magstep1

\font\bfXII=cmbx10 scaled\magstep1

\def\sa{\fam10\msamX}
\font\msamX=msam10

\def\carre{\hbox{\sa\char'004}}
\def\bl{\leavevmode\hbox{$\bullet$}}
\def\d{\delta}

\gauche={Serge Burckel}
\droite={Non-Interlaced SAT is in P.}
\def\P{{\bf P}}
\def\Z{{\bf Z}}
\def\NP{{\bf NP}}

\def\DT{{\bf DTIME }}

\def\x{{\hskip -3pt}+{\hskip -3pt}}
\def\xx{+}
\def\p1{\x 1}
\def\xx10{\x 1_0}

\def\et{\wedge}
\def\ou{\vee}

\def\ex{\exists}

\def\x{{\hskip -2pt}+{\hskip -2pt}}

\def\et{{\hskip -2.5pt}\wedge{\hskip -2.5pt}}

\def\ffll{\rightarrow}

\def\fl{{\hskip -2.5pt}\ffll{\hskip -2.5pt}}

\def\hfl#1#2{\smash{\mathop{\hbox to 5truemm{\rightarrowfill}}  \limits^{\scriptstyle#1}_{\scriptstyle#2}}}

\def\hfl#1{\smash{\mathop{\hbox to 8truemm{\rightarrowfill}}  \limits^{\scriptstyle#1}}}

\def\Rem#1{}
\def\et{\wedge}
\def\ou{\vee}
\def\a{\alpha}
\def\b{\beta}
\def\c{\gamma}
\def\T{\Gamma}
\def\D{\Delta}

\centerline{\XII Non-Interlaced SAT is in \P.}
\bn
\centerline{\sl Dr Serge Burckel, S.I.G.L.E. (www.sigle.space).}
\bn
{\bf Abstract. We investigate the \NP-Complete problem SAT and the geometry of its instances. For a particular type that we call {\it non-interlaced formulas}, we propose a polynomial time algorithm for their resolution using graphs and matrices. }
\bn\ttparag{1. Introduction.}
\sk We investigate the decision problem SAT. Some reductions have been done in order to preserve the property to be \NP-Complete ([1], [2]). Here, we propose a geometrical property that depends on the order of the clauses and that can lead to a polynomial time method. We consider {\it formulas} represented by finite lists $F=[C_1,C_2,\dots,C_k]$ where each {\it clause}~$C_i$ is a finite list $(x^i_1,\dots,x^i_{n_i})$ where $n_i>0$ and each $x^i_a\in \Z^*$. This is an easy way to encode general CNF instances. For example $(a\ou b\ou\lnot c)\et(\lnot a\ou\lnot b)$ will be encoded by $[(1,2,-3),(-1,-2)]$. This is even more general because we allow clauses like $(1,2,1,-2,-1)$. The SAT problem is to decide whether there exists a {\it good choice} $[c_1,\dots,c_k]$ where each $c_i\in C_i$ and $c_i\not=-c_j$ for every~$i,j$. 
\bn We will count the total number $\T(F)$ of these good choices since the initial formula $F$ is satisfiable if and only if $\T(F)>0$. This will be possible in polynomial time for a certain type of instances.
\bn\ttparag{2. Non-Interlaced Formulas.}
\Def{ Let $F$ be a formula.
\sk Let $\D(F)$ be the set of pairs $[i,j]$ where $i<j$ and $\ex x^i_a,x^j_b$ with $x^i_a=-x^j_b$. We say that $F$ is {\it interlaced} if $\ex[i,j],[i',j']\in \D(F)$ with $i<i'<j<j'$. Otherwise, $F$ is {\it non-interlaced}.}
\bn For example $F=[(1),(2),(-1),(-2)]$ is interlaced with $\D(F)=\{[1,3],[2,4]\}$ and the permuted formula $F'=[(1),(2),(-2),(-1)]$ is non-interlaced with $\D(F')=\{[1,4],[2,3]\}$. 
\bn We call {\it Non-Interlaced SAT} the restriction of the SAT problem to non-interlaced formulas.
\bn \Thm{ Non-Interlaced SAT is in \P. }
\bn The proof consists of building an effective polynomial time algorithm that counts the number $\T(F)$ of good choices with a graph $G$ and an associated matrix $M$.
\bn\ttparag{3. The Graph and the Matrix.}
The graph $G$ has $N=2+n_1+\dots n_k$ vertices~:  two special vertices $s,t$ and  a vertex $v^i_a$ for each~$x^i_a$. The edges of $G$ are directed and valued. Moreover, they are of different types in order to distinguish them. We do not indicate this type when it is not necessary. However, notice that two vertices can be linked with edges of different types. We begin to put the edges of {\it type 0} in $G$~:
\sk $s\hfl 1 v^1_a$ for every $x^1_a\in C_1$
\sk $v^i_a\hfl 1 v^j_b$ for every $1\le i<k$ and $j=i+1$
\sk $v^k_a\hfl 1 t$ for every $x^k_a\in C_k$
\sk $s\hfl 1 t$ when $k=0$
\sk Observe that $G$ has no cycle. This property will be preserved in the sequel.
\bn In practice, we will not have to build $G$. We will build simultaneously its adjacency matrix~$M$ of dimension $N\times N$ indexed by the vertices of $G$ and such that $M[x,y]$ is the sum of the values of the edges $x\fl y$.
\hfill\eject
\Def{ For two vertices $x,y$, the number $\pi(x,y)$ is the value of all the paths from $x$ to $y$ which is equal to the sum over all these paths of the products of the values of their edges.}
\bn In order to compute $\pi(x,y)$, we perform the following operations on the matrix~$M$ since the longest path will always have at most $k+1$ edges~:
\bn{\tt Fix $M[y,y]:=1$. Compute $M':=M^{k+1}$.  Fix $M[y,y]:=0$. Return the entry $M'[x,y]$.}
\bn Observe that fixing $M[y,y]:=1$ enables us to take account of all the paths whatever their lengths.
\bn Observe that $k+1<N$. Hence, this computation can obviously be performed in $O(N^4)$ steps.
\bn\ttparag{4. New edges.}
We expect to have $\pi(s,t)=\T(F)$. We are going to ``remove'' step by step the contributions of pairs $[v^i_a,v^j_b]$ where $x^i_a=-x^j_b$. We consider these pairs in the increasing order of $(j-i)$.
\bn{\tt
For $\d$ from $1$ to $k-1$ do
\sk\hskip 4mm For every pair $[i,j]\in \D(F)$ with $j=i+\d$
\sk\hskip 8mm  For every pair $[v^i_a,v^j_b]$ with $x^i_a=-x^j_b$
\sk\hskip 12mm  Compute the number $\a:=\pi(v^i_a,v^j_b)$.
\sk\hskip 12mm  Add an edge $v^i_a\hfl{-\a}v^j_b$ of type $1$.
} 
\bn Observe for example, that an edge of type $0$ like $v^i_a\hfl 1 v^{i+1}_b$ with $x^i_a=-x^{i+1}_b$ leads to the creation of an edge of type $1$~: $v^i_a\hfl {-1} v^{i+1}_b$. Hence $M[v^i_a,v^{i+1}_b]=0$. We have the complete method~:
\bn{\tt
1. Given a non-interlaced formula $F$, compute the set $\D(F)$.
\sk 2. Build the matrix $M$ according to edges of type $0$.
\sk 3. Add the edges of type $1$ in $M$.
\sk 4. Conclude that $F$ is satisfiable if and only if $\pi(s,t)>0$.
}
\bn We are able to prove Theorem $1$.
\sk\Pr\ We show that the previous method gives in polynomial time a correct answer.
\sk\bl\ {\bf Complexity.} The number $N$ is smaller than the size of the instance $F$. Since the number of pairs~$[v^i_a,v^j_b]$ is bounded by $N^2$, the algorithm is polynomial in $\DT(N^6)$. 
\sk The computed numbers like $\T(F)$ are all bounded by the product $n_1.n_2\dots n_k$ and representable in space $log(n_1)+log(n_2)+\dots log(n_k)<N$.
\sk\bl\ {\bf Correctness.} An edge $e$ of type $1$ like $v^i_a\hfl{-\a}v^j_b$ will set to zero the values of all the paths containing these two vertices. By induction on $\d=j-i$ that is sufficient to be correct. For $\d=1$, that is obvious and we have already made this observation. For $\d>1$, the possible edges of type $1$ behind this edge $e$ have made (by induction hypothesis) the good corrections for $\pi(v^i_a,v^j_b)$ that will be the number of remaining good paths between $v^i_a$ and $v^j_b$. The edge $e$ removes them with a value $-\a$. For the others edges $f$ of type $1$ with values $-\b$ that are before of after $e$, the inclusion-exclusion principle is automatically applied with the structure of the graph. The paths from $s$ to $t$ have $4$ possibilities~:
\sk not use $e$ and not use $f$~: the non corrected paths,
\sk use $e$ and not use $f$~: minus the ones corrected by $e$~:$-\a$,
\sk not use $e$ and use $f$~: minus the ones corrected by $f$~:$-\b$,
\sk use $e$ and use $f$~: plus the ones corrected by $e$ and by $f$ (that where removed twice)~:$(-\a)(-\b)=\a\b$.
\sk The same inclusion-exclusion principle applies for any number of successive edges of type $1$. \Rp
\bn\ttparag{5. Conclusion.}
When the formula $F$ is non-interlaced, we always have $\pi(s,t)=\T(F)$. 
\sk However, there exist interlaced formulas for which this is not the case. 
\sk For example, the computation for $F=[(1),(2),(-1),(-2)]$ will give $\pi(s,t)=-1$ whereas $\T(F)=0$.
\bn We hope that this paper will bring some interests in the investigation of the geometry of formulas and lead to future improvements. One could adapt the presented method for interlaced formulas or find a tricky polynomial time transformation $\Phi$ such that an interlaced formula $F$ becomes a non-interlaced formula $\Phi(F)$ that is satisfiable if and only if $F$ is. That would imply $\P=\NP$.

\bn\bn

[1] M.R. Garey, D.S. Johnson, Computers and Intractability, Freeman, New York (1979).

[2] C.A. Tovey, A simplified satisfiability problem, Discrete Appl. Math., 8 (1984), pp. 85-89.

\vfill
\end

\bn\ttparag{5. What about interlaced formulas ?}
\bn According to what we have done previously, it is quite natural to introduce new edges. These edges will be of type $2$ and perform the convenient corrections for interlaced edges of type $1$. These corrections where not performed automatically with products that correspond to the inclusion-exclusion principle because of this interweaving. We do that by hand with a new procedure~:
\bn{\tt
\sk For $t=1$
\sk\hskip 4mm For $\d$ from $1$ to $k-1$ do
\sk\hskip 8mm For every pairs of edges of type $t$~: $v^i_a\fl v^j_b$ and $v^h_c\fl v^k_d$ 
\sk\hskip 8mm if $i<h<j<k$ and $k=i+\d$ then
\sk\hskip 12mm  Remember and remove the outgoing edges of the $v^j_{b'},v^h_{c'}$ with $b'\not=b$ and $c'\not=c$.
\sk\hskip 12mm  Compute the number $\a:=\pi(v^i_a,v^k_d)$.
\sk\hskip 12mm  Add an edge $v^i_a\hfl{\c}v^k_d$ of type $t+1$ where $\c=(-1)^{t+1}\a$
\sk\hskip 12mm  Restore the remembered edges.
} 
\bn That will perform the corrections on these edges of type $1$~: for two interlaced edges $e,f$ of type~$1$, we count the paths from $v^i_a$ to $v^k_d$ that use all their vertices because we have removed the other possibilities. Now, we can do the same for interlaced edges of type $2,3,\dots$
\sk We just have to replace {\tt For $t=1$} by {\tt For $t=1,2,\dots K$}. We obtain by this way a polynomial time algorithm for SAT since the minimal length of the edges of type $t$ increases with $t$.
\bn \Thmm{ SAT is in \P. }